\documentclass[aps,prd,amsfonts,amssymb,preprint,eqsecnum,nofootinbib]
{revtex4} 
\usepackage{graphics,epsfig}
\begin{document}
\preprint{WM-05-114}
%
\title{\vspace*{0.5in} Five-dimensional Trinification Improved \vskip
0.1in}
\author{Christopher D. Carone}\email[]{carone@physics.wm.edu}
\author{Justin M. Conroy}\email[]{jmconr@wm.edu}
\affiliation{Particle Theory Group, Department of Physics,
College of William and Mary, Williamsburg, VA 23187-8795}
\date{July 2005}

\begin{abstract}
We present improved models of trinification in five dimensions. Unified symmetry is
broken by a combination of orbifold projections and a boundary Higgs sector.  The latter
can be decoupled from the theory, realizing a Higgsless limit in which the scale of exotic
massive gauge fields is set by the compactification radius.  Electroweak Higgs doublets
are identified with the fifth components of gauge fields and Yukawa interactions arise via
Wilson loops.  The result is a simple low-energy effective theory that is consistent with
the constraints from proton decay and gauge unification.
\end{abstract}
\maketitle

\section{Introduction}\label{sec:intro}

Extra dimensions provide a variety of new tools for building realistic Grand
Unified Theories (GUTs).  In orbifold compactifications, for example, different
components of a GUT multiplet may be assigned different parities under reflections
about the orbifold fixed points.  Judicious choices can yield a particle spectrum in
which all unwanted states (for example, color-triplet Higgs fields) appear at or
near the compactification scale $1/R$.  A related technique that has received some
attention in the context of electroweak symmetry breaking is the Higgsless
mechanism~\cite{higgsless,morehiggsless}.  In this approach, a more general set of boundary
conditions are employed, allowing for the reduction in the rank of the gauge group.  These
boundary conditions can be thought of as arising from a boundary Higgs sector that
has been decoupled from the theory.  Interestingly, in this decoupling limit, the
spectrum of massive gauge fields is determined by $1/R$ rather than the boundary
vacuum expectation values (vevs)~\cite{nomura}.  While electroweak symmetry breaking
clearly necessitates the reduction in rank of the gauge group, the same is true of
GUTs with rank greater than four.  This was the motivation for the study of
boundary breaking in trinified theories~\cite{carcon}, one of the simplest unified
theories of rank six.  Other recent work on trinified theories in extra dimensions appears
in Ref.~\cite{othertrin}.

While Ref.~\cite{carcon} explored the usefulness of generalized boundary
conditions in breaking a simple unified theory of rank greater than four, the
models presented there had a number of shortcomings:  electroweak symmetry breaking
was still accomplished by introducing chiral Higgs multiplets and a fine-tuning
was required to keep these fields in the low-energy spectrum.  In this letter, we
present simpler models that avoid these problems.  Electroweak Higgs doublets will
be identified as components of gauge fields, an economical approach
known as gauge-Higgs unification in the literature~\cite{gh,gh2}, and these Higgs fields will
remain light down to the weak scale due to an R-symmetry~\cite{nomurahall}.  In addition, we
present one construction in which an additional gauge group factor provides
both for a unified boundary condition on the standard model gauge couplings and
also serves as an origin for the electroweak Higgs fields.  This yields a trinified
theory without the cumbersome (though entirely conventional) cyclic symmetry whose only
purpose is to  maintain the equality of GUT-scale gauge couplings.  The two models we
present are consistent with the constraints from proton decay and gauge coupling
unification.

\section{$SU(3)^3\ltimes \mathbf{Z}_3$}

Conventional trinification is based on the gauge group $G_T=SU(3)_C \times SU(3)_L
\times SU(3)_R \ltimes Z_3$.  The discrete symmetry cyclically permutes the group
labels C,L, and R, which maintains a single gauge coupling $g$ at the unification scale.
Gauge and matter fields transform under the ${\bf 24}$- and ${\bf 27}$-dimensional
representations, respectively, with decompositions
\begin{eqnarray}
\mathbf{24} &=& ({\bf 8},{\bf 1},{\bf 1})\oplus ({\bf 1},{\bf 8}, {\bf 1})\oplus
({\bf 1},{\bf 1},{\bf 8})
\,\,\, , \nonumber \\
\mathbf{27} & = & (\mathbf{1,3,\bar{3}}) \oplus (\mathbf{\bar{3},1,3})
\oplus (\mathbf{3,\bar{3},1}) \,\,\,,
\end{eqnarray}
under the C, L, R gauge factors.  In the usual Gell-Mann basis, weak SU(2) is generated
by $T^a_L$ for $a=1\ldots 3$, while hypercharge, in its standard model normalization,
is generated by
\begin{equation}
Y = -\frac{1}{\sqrt{3}} ( T^8_L + \sqrt{3} T^3_R + T^8_R ) \,\,\, .
\label{eq:hypgen}
\end{equation}
With the hypercharge gauge coupling identified as $\sqrt{3/5} g$, the choices
above yield the standard GUT-scale prediction $\sin^2{\theta_W}=3/8$.  This is
phenomenologically acceptable in the present context, given the new boundary
corrections to unification~\cite{nomura} that we expect generically in extra-dimensional
models.

We first consider a model in five dimensions (5D) with $G_T$ chosen as the
bulk gauge symmetry.  We compactify the extra dimension on an $S^1/(Z_2 \times Z_{2}')$
orbifold, labelled by the coordinate $y$.  Defining $y'\equiv y+\pi R/2$, points related
by the translation $y\rightarrow y + 2 \pi R$ and by the reflections $y \rightarrow -y$ and
$y'\rightarrow -y'$, are identified.   The physical region in $y$ is thus reduced to the
interval $[0,\pi R/2]$.  In addition, we assume ${\cal N}=1$ supersymmetry in 5D.  Bulk
gauge fields thus form ${\cal N}=2$ 4D hypermultiplets consisting of ${\cal N}=1$ vector
$V(A^{\mu},\lambda)$ and chiral $\Phi(\sigma+i A_5,\lambda')$ multiplets at each
Kaluza-Klein (KK) level.  All matter fields are placed on the $\pi R/2$ brane for simplicity.

We now show that the electroweak Higgs doublets of the minimal supersymmetric
standard model (MSSM) can be identified with some of the $A_5$ components of
the gauge multiplets.  Under the two orbifold parities, we assume the bulk fields
transform as follows:
\begin{eqnarray}
V(x^\mu,-y) &=& P\,V(x^\mu,y)\, P^{-1} \,, \,\,\,\,\,\,\,\,\,\,
V(x^\mu,-y') = P'\,V(x^\mu,y')\, P^{\prime -1} \nonumber \\
\Phi(x^\mu,-y) &=& -P\,\Phi(x^\mu,y)\, P^{-1} \, , \,\,\,\,\,\,\,
\Phi(x^\mu,-y') = -P'\,\Phi(x^\mu,y')\, P^{\prime -1}  \,\,\,.
\label{eq:vphi}
\end{eqnarray}
Here $P$ and $P'$ are $3 \times 3$ matrices that act in gauge group
space and have eigenvalues of $\pm 1$.  Noting that the supersymmetric
bulk action requires the terms $\mathcal{S}_{5D} \supset \int d^4 \theta
\frac{2}{g^2}Tr(\sqrt{2}\partial_5
+\Phi^{\dag})e^{-V}(-\sqrt{2}\partial_5 +\Phi)e^V$~\cite{gregoire},
one sees that $\partial_5 V$ and $\Phi$ should have the same
transformation properties under the orbifold parities. Therefore, although
components within a gauge multiplet can transform differently under the
parity operations, the relative sign of the vector and chiral multiplets
is uniquely determined. With the notation $(P,P')=(P_C \oplus P_L \oplus P_R, P'_C \oplus
P'_L \oplus P'_R$) we choose
\begin{eqnarray}
P_C &=& diag(1,1,1), \,\,\,\,\, P_L =diag(1,1,-1), \,\,\,\,\, P_R
=diag(1,1,-1), \nonumber \\
P'_C &=& diag(1,1,1), \,\,\,\,\, P'_L =diag(1,1,-1), \,\,\,\,\, P'_R =diag(1,1,1).
\end{eqnarray}
Parity assignments for the component fields immediately follow:
\begin{equation}
V_C:\: \left(
\begin{array}{cc|c} (+,+) & (+,+) & (+,+) \\ (+,+) & (+,+) & (+,+)
\\ \hline (+,+) & (+,+) & (+,+) \end{array} \right), \qquad
\Phi_C:\: \left( \begin{array}{cc|c} (-,-) & (-,-) & (-,-) \\
(-,-) & (-,-) & (-,-) \\ \hline (-,-) & (-,-) & (-,-) \end{array}
\right),  \end{equation}
\begin{equation} V_L:\: \left( \begin{array}{cc|c} (+,+) & (+,+) &
(-,-) \\ (+,+) & (+,+) & (-,-) \\ \hline (-,-) & (-,-) & (+,+)
\end{array} \right), \qquad \Phi_L:\: \left( \begin{array}{cc|c}
(-,-) & (-,-) & (+,+) \\ (-,-) & (-,-) & (+,+) \\ \hline (+,+) &
(+,+) & (-,-) \end{array} \right),
\end{equation} \begin{equation} V_R:\: \left( \begin{array}{cc|c}
(+,+) & (+,+) & (-,+) \\ (+,+) & (+,+) & (-,+) \\ \hline (-,+) &
(-,+) & (+,+) \end{array} \right), \qquad \Phi_R:\: \left(
\begin{array}{cc|c} (-,-) & (-,-) & (+,-) \\ (-,-) & (-,-) & (+,-)
\\ \hline (+,-) & (+,-) & (-,-) \end{array} \right).
 \end{equation}
As we will see shortly, fields that are odd under $P$ have vanishing
wave functions at $y=0$, while those that are odd under $P'$ vanish
at $y=\pi R/2$.  It follows that the gauge symmetry that is operative
at the $\pi R/2$ fixed point is $SU(3)_C\times SU(2)_L \times U(1)_L \times
SU(3)_R$, a fact that we will use later. Only fields that are even under both $P$
and $P'$ have massless zero modes, from which we conclude that the total effect
of the orbifold projection is to reduce the bulk gauge symmetry to
$SU(3)_C \times SU(2)_L \times U(1)_L \times SU(2)_R \times U(1)_R$.
Crucially, two SU(2)$_L$ doublets in the chiral multiplet $\Phi_L$
retain massless zero modes, and it follows immediately from
Eq.~(\ref{eq:hypgen}) that these have hypercharges $Y=\pm \frac{1}{2}$.  We
identify these superfields with the MSSM Higgs doublets.

We break the remaining gauge symmetry down to that of the MSSM
using generalized boundary conditions. To illustrate
this approach consider a gauge field $A^\mu$ that is
even under reflections about $y=0$.  This implies that the
5D wave function for the $k^{th}$ mode has the form
\begin{equation}
A_{\mu} (x^{\nu},y)\sim \cos(M_k\, y)A^{(k)}_{\mu}(x^{\nu}) \,\,\,,
\end{equation}
for $y$ in the interval $0\leq y \leq \pi R/2$. Imposing the boundary condition
\begin{equation}
\partial_5 A^{\mu}(y=\pi R/2)=V A^{\mu}(y=\pi R/2)\,,
\label{eq:bc}
\end{equation}
one obtains the following transcendental equation for $M_k$
\begin{equation} M_{k}
\tan(M_{k}\pi R/2)=-V \,\,\,.
\end{equation}
In the large V limit the KK spectrum is well approximated by
\begin{equation} M_k \approx
M_{c}\frac{(2k+1)}{2}(1+\frac{M_c}{\pi V}+\cdots), \,\,\,\,\, k=0,1,\ldots \,,
\label{eq:transc1}
\end{equation}
where we define the compactification scale $M_{c}\equiv 2/R$.  Thus, in the limit
$V \rightarrow \infty$, the spectrum reduces to a tower whose low lying states are
$M_c/2$, $3M_c/2$, $5M_c/2$, {\em etc}. This is shifted by $M_c/2$ relative to the tower
one would obtain if $V$ were set to zero. The symmetry breaking parameter V has dimensions
of mass and can be associated with products of the form $g^2 v^2$, where $g$ is a
five-dimensional gauge coupling and $v$ a boundary vev.  Since $v$ generically sets
the scale of the physical states in the boundary symmetry-breaking sector, the limit
$V \rightarrow \infty$ corresponds to the decoupling of the boundary Higgs fields
from the theory.  It is worth noting that in the supersymmetric case, the
spectrum of the additional scalar and fermionic components of $\Phi$ and $V$ are the
same as in Eq.~(\ref{eq:transc1}), as a consequence of gauge invariance and unbroken
supersymmetry~\cite{nomura}.

In the present context, we could introduce two ${\bf 27}$ boundary Higgs fields,
whose $({\bf 1}, {\bf 3}, {\bf \bar{3}})$ components have appropriate vevs to
break SU(3)$^3$ down to the standard model gauge group.  However, we have already
noted that the orbifold projection has reduced the gauge symmetry to $SU(3)_C \times
SU(2)_L \times U(1)_L \times SU(3)_R$ at the $\pi R/2$ brane.  This allows us to
choose much simpler Higgs representations at this boundary to implement
the symmetry breaking:
\begin{equation} \chi_1=\bar{\chi}_1= \pmatrix{0 \cr 0 \cr v_1} , \,\,\,\,\, \chi_2=\bar{\chi}_2
= \pmatrix{0 \cr v_2 \cr v_3} \,.
\label{eq:trips}
\end{equation}
Here the field $\chi$ is an SU(3)$_R$ triplet with U(1)$_L$ charge $+1/\sqrt{3}$,
and $\bar{\chi}$ has conjugate quantum numbers.  Both $\chi_i$ and $\bar{\chi}_i$ are
singlets under color and $SU(2)_L$, and are together anomaly free.  They represent the one
relevant row of the $({\bf 1}, {\bf 3}, {\bf \bar{3}})$ representation used to break the
unified symmetry in conventional trinified models.  The real vevs shown in $\chi_1$ and
$\chi_2$ are completely general choices, while we assume the same vevs for the barred fields.
(The latter choice is consistent with $D$-flatness.)  These vevs are sufficient to break the
remaining gauge symmetry down to that of the standard model.  Note, however, that a realistic
potential may require additional fields.

How do these boundary vevs affect the spectra of fields transforming as $(+,+)$, $(-,-)$,
$(-,+)$, and $(+,-)$ under our orbifold reflections? First, the fields $A^\mu(x^\nu,y)$ whose
wave functions are odd at the $y=\pi R/2$ brane (corresponding to parities
$(\pm, -)$),  vanish at that endpoint.  Therefore, these KK towers are
unaffected by the boundary vevs, and are given by $M_{n}(-,-)=(n+1)M_c$ and
$M_{n}(+,-)=(n+1/2)M_c$, for $n=0,1,\ldots$.  As discussed above, the $(+,+)$
fields acquire a massive tower $M_n(+,+,V\rightarrow\infty)= (n+1/2)M_c$.  Finally,
consider the $(-,+)$ fields. Since these  wave functions are odd at the $y=0$ brane
they have the general form
\begin{equation}
A_{\mu}(x^{\nu},y)\sim \sin(M_k\, y)A_{\mu}^{(k)}(x^{\nu}) \,\,\,.
\end{equation}
Imposing the boundary condition in Eq.~(\ref{eq:bc}) yields the transcendental equation
\begin{equation}
M_{k} \cot(M_k \pi R/2)= V \,\,\,,
\end{equation}
which implies a KK tower
\begin{equation} M_k \approx
M_{c}\,k\, (1+\frac{M_c}{\pi V}+\cdots)\,, \,\,\,\,\, k=1,2,\ldots \,,
\end{equation} in the large $V$ limit.

In the present case, the boundary conditions following from the existence of the
$\chi$ and $\bar{\chi}$ vacuum expectation values may be written
\begin{equation}
\partial_5 A^i_\mu(x^\nu,\pi R/2) = V_{ij} A^j_\mu(x^\nu, \pi R/2)\,\,\, ,
\end{equation}
where $V_{ij}$ is a matrix in the space of the SU(3)$_L\times$SU(3)$_R$ gauge fields.
The entries of this matrix were considered explicitly in Ref.~\cite{carcon}, and
have the form $\sum_i c_i g^2 v_i^2$, where the $v_i$ are defined in Eq.~(\ref{eq:trips})
and the $c_i$ are numerical coefficients.  The precise form of $V_{ij}$ and the values of
the $c_i$ are irrelevant for the present analysis since we will always take the $g^2 v^2_i$
to be large compared to $M_c$. In this limit, the spectra of KK modes become independent
of these details, and we obtain one of six possible towers already discussed: $(+,-)$, $(-,-)$,
$(+,+,V=0)$, $(+,+,V\rightarrow\infty)$, $(-,+,V=0)$, and  $(-,+,V\rightarrow\infty)$.

As noted earlier, matter fields are located at the $\pi R/2$ fixed point.  Although
the SU(3)$^3$ ${\bf 27}$s decompose into a direct sum of representations under the
unbroken gauge symmetry at $\pi R/2$, all of these components must be retained in order
to have an anomaly-free theory that reproduces complete MSSM generations in the low-energy
theory.  We now show that the exotic matter content of the ${\bf 27}$s become massive
via couplings to the boundary Higgs fields $\chi$ and $\bar{\chi}$, and decouple from
the theory as the vevs $v_i$ are taken large.  We first decompose the ${\bf 27}$
under the unbroken SU(3)$_C\times$SU(2)$_L \times$U(1)$_L \times$ SU(3)$_R$ symmetry
at $y=\pi R/2$,
\begin{equation}
{\bf 27}= L({\bf 1},{\bf 2},{\bf \bar{3}})_q+e^c({\bf 1},{\bf 1},{\bf \bar{3}})_{-2q}+
q^c({\bf \bar{3}},{\bf 1},{\bf 3})_0 + Q({\bf 3},{\bf 2},{\bf 1})_{-q} +
B({\bf 3},{\bf 1},{\bf 1})_{2q}\,\,\,,
\label{eq:twentyseven}
\end{equation}
where the U(1)$_L$ charge $q=1/(2\sqrt{3})$. The notation serves as a reminder of the
embedding of standard model fields. In addition, the $L$ multiplet contains a vector-like
pair of exotic lepton doublets, $(E^0, E^-)$ and $(E^+, E^{0c})$, $e^c$ contains a pair of
exotic singlets, $N^c$ and $N$, while $q^c$ contains the right-handed partners of the
exotic left-handed, charge $-1/3$ quarks that make up the multiplet $B$ in its entirety.
GUT scale mass terms arise via the superpotential couplings
\begin{equation}
W =h_L (L^{i}_{\alpha}L^{j}_{\beta}
\bar{\chi}_{\gamma} \epsilon^{\alpha \beta \gamma} \epsilon_{ij})+
\frac{1}{2}h_Q({q^c}_i^{\alpha}B^{j}\bar{\chi}_{\beta}\delta_{j}^{i}\delta_{\alpha}^{\beta})\,.
\label{eq:gutmass}
\end{equation}
Expanding Eq.(\ref{eq:gutmass}) produces the low-energy matter content of minimal trinified
theories. The right-handed $d$ and $B$ quarks mix leaving one linear combination
massless to be identified with the physical $d_R$ quark.  Similarly, only one linear
combination of the left-handed lepton doublets receives a mass from the first term in
Eq.~(\ref{eq:gutmass}).  Thus, the low-energy spectrum consists of the particle content
of the MSSM, as well as the singlets $N$ and $N^c$.   These can be made massive
as well by including higher-dimension operators in the superpotential of the form,
$(e^c \chi_i)^2/\Lambda$, where $\Lambda$ is the cutoff of the effective theory.   Thus,
unlike the model in Ref.~\cite{carcon}, no additional fields need to be included at
the boundary to rid the low-energy theory of the singlets.

Having recovered the MSSM particle content, we now consider how to obtain Yukawa
couplings involving the electroweak Higgs doublets. Since we have identified
the Higgs doublets with components of the bulk adjoint chiral superfield $\Phi$,
which transforms nonlinearly under a gauge transformation ({\em i.e.}.
$\Phi \rightarrow e^{\Lambda}(\Phi-\sqrt{2}\partial_5) e^{-\Lambda}$.), no
local Yukawa couplings are possible. However, the solution to this problem is
well known in the literature on 5D gauge-Higgs unification models: one may couple
the Higgs doublets to the matter fields at the fixed point via Wilson loop
operators~\cite{gh2,nomurahall}. The Wilson line operator
${\cal H}=\mathcal{P}exp(\int_{y_i}^{y_f}\frac{1}{\sqrt{2}}\Phi dy)$,
where $\mathcal{P}$ represents the path ordered product, is a
nonlocal object that transforms linearly under the 5D gauge
transformation at points $y_i$ and $y_f$, ${\cal H} \rightarrow e^{\Lambda}|_{y_f} {\cal H}
e^{-\Lambda}|_{y_i}$. Choosing $y_i=y_f=\pi R/2$ and a path that wraps around the extra
dimension, one obtains a Wilson loop operator that transforms linearly at the orbifold
fixed point where our matter fields are located: ${\cal H} \rightarrow
e^{\Lambda} {\cal H} e^{-\Lambda}$, with $\Lambda \equiv \Lambda(x^\mu,y=\pi R/2)$.
We focus on the doublet components of ${\cal H}$, which we call
$H({\bf 1},{\bf 2},{\bf 1})_{3q}$ and $\bar{H}({\bf 1},{\bf 2},{\bf 1})_{-3q}$, using
the notation of Eq.~(\ref{eq:twentyseven}).  Yukawa couplings originate at the $\pi R/2$
brane via the interactions
\begin{equation}
W = \frac{1}{\Lambda} \bar{\chi} L e^c H +
\frac{1}{\Lambda} \bar{\chi} Q q^c H +
\frac{1}{\Lambda^2} \chi_1 \chi_2 Q q^c \bar{H} \,\,\, ,
\label{eq:yukawas}
\end{equation}
after the $\chi$ and $\overline{\chi}$ fields develop vevs.  Here $\Lambda$
is a cutoff of the effective theory.   Note that in the decoupling limit
$v_i \rightarrow \Lambda \rightarrow\infty$, none of the terms in Eq.~(\ref{eq:yukawas})
are suppressed; this is an indication that the low-energy theory is restricted only by
standard model gauge symmetry at $y=\pi R/2$.

We resolve the $\mu$ problem in our model by using the $U(1)_R$ symmetry of the bulk
action.  Under this symmetry, the superspace coordinate $\theta$ transforms with
charge $+1$, while $V$ and $\Phi$ are neutral.  An $H\bar{H}$ term is not allowed
since the superpotential must have $R$-charge $-2$. We may induce a small $\mu$
parameter by coupling the Higgs fields to a singlet $X$ with $R$-charge $-2$, via
the superpotential coupling $X H \bar{H}$.  The $\mu$ parameter is generated if
the $X$ field develops a vev, which can happen naturally due to supersymmetry-breaking
effects, as in the next-to-minimal supersymmetric standard model.  Note that
this mechanism works assuming we impose only a discrete subgroup of U(1)$_R$, which
avoids any unwanted $R$-axions.  Assuming the $\chi$ and $\bar{\chi}$ have $R$-charge
zero and each matter field $-1$, then the Yukawa couplings in Eq.~(\ref{eq:yukawas})
are allowed and a $Z_4$ subgroup of U(1)$_R$ is sufficient.

Finally, we consider the issue of gauge coupling unification.  The possible towers
of KK modes are described by either $(n+1/2)M_c>0$ or $n M_c>0$, for $n$ an integer.
The supersymmetric beta functions for the fields charged under the standard model
gauge groups are shown in Table~\ref{betatable}. Note that only two exotic
$(V,\Phi)$ multiplets, with charges $({\bf 1},{\bf 1},{\bf 1})$ and $({\bf 1},{\bf 1},{\bf -1})$,
respectively, under SU(3)$_C\times$ SU(2)$_W\times$U(1)$_Y$ have KK towers that are shifted
down by $M_c/2$ due to the boundary Higgs vevs.
\begin{table} \begin{center}
\begin{tabular}{ccccc}\hline\hline &$(b_3,b_2,b_1)$ \qquad &
\qquad ($\tilde{b}_3,\tilde{b}_2,\tilde{b}_1$) \\ \hline
$(V,\Phi)_{321} $  &  (-9,-6,0) \qquad & \qquad (-6,-4,0) \\
$(V,\Phi)_{(1,2,\pm 1/2)}$  &   (0,1,$\frac{3}{5}$) \qquad & \qquad
(0,-2,-$\frac{6}{5}$) \\ $(V,\Phi)_{(1,1,\pm 1)} $ &    - \qquad &
\qquad (0,0,$-\frac{24}{5}$) \\ Matter & (6,6,6) \qquad & \qquad
-  \\ \hline Total & (-3,1,$\frac{33}{5}$) \qquad & \qquad
(-6,-6,-6)  \\ \hline\hline \end{tabular}
\end{center}
\caption{Contributions to the beta function coefficients from the
zero modes ($b_i$) and the KK levels ($\tilde{b}_i$).  Here the $\Phi$ represent
chiral multiplets in the adjoint representation.  Results in the second and
third lines represent sums over all fields with the stated quantum numbers.}\label{betatable}
\end{table}
Notice that if all the KK towers were aligned, they would contribute universally
to the gauge running.  As first pointed out in Ref.~\cite{nomura}, the shifted spectra
contribute as a tower of threshold corrections;  while the power-law running is still
universal, the running of the differences $\alpha_{i}^{-1}-\alpha_{j}^{-1}$ is logarithmic.
With the notation $\delta_{i}(\mu)=\alpha_{i}^{-1}(\mu)-\alpha_{1}^{-1}(\mu)$, for
$i=2\mbox{ or }3$, the differential running above the first KK threshold is given by
\begin{equation}
\delta_{i}(\mu)=\delta_{i}(M_c/2)-\frac{1}{2\pi} R_{i}(\mu) \,\,\,,
\end{equation}
where
\begin{equation}
R_2(\mu)=-\frac{28}{5} \log ( \frac{\mu}{M_c/2})- \frac{12}{5} \sum_{0<nM_c<\mu}\log(
\frac{\mu}{nM_c})+\frac{12}{5}\!\sum_{0<(n+1/2)M_c<\mu}\!\log(\frac{\mu}{[n+1/2]M_c}),
\end{equation}
\begin{equation}
R_3(\mu)=-\frac{48}{5} \log ( \frac{\mu}{M_c/2})- \frac{12}{5}\sum_{0<nM_c<\mu}\log(
\frac{\mu}{nM_c})+\frac{12}{5}\!\sum_{0<(n+1/2)M_c<\mu}\!\log(\frac{\mu}{[n+1/2]M_c})\,.
\end{equation}
Note that the last two terms in each equation above would cancel if the KK-towers were aligned,
and one would obtain the differential running of the MSSM.  Numerical study of these
equations reveal that unification is preserved, but that the scale of unification $M_U$ is delayed.
For example, for $M_C = 4\times 10^{14}$~GeV we find $M_U \approx 8\times 10^{16}$~GeV, which is
approximately the 5D Planck scale.  For $M_C=2\times 10^{16}$~GeV we find
$M_U=2.8 \times 10^{16}$~GeV, which simply demonstrates that there is a limit in
which most of the KK towers do not contribute and MSSM unification is recovered.  For
$4\times 10^{14}$~GeV$<M_U<2\times 10^{16}$~GeV we find that the $\alpha^{-1}$ unify
at well below the 1\% level, ignoring possible boundary effects.  Thus, our extra-dimensional
construction does not lead to any problems with successful gauge unification.  Discussion of
other possible corrections to unification may be found in Ref.~\cite{carcon} and will not be
discussed further here.

Finally, we note that there is no proton decay in this model.  In ordinary trinification,
proton decay is mediated by colored Higgses that are part of a ${\bf 27}$.  In our model,
the smaller gauge symmetry at the $\pi R/2$ fixed point allowed us to include symmetry
breaking fields in much smaller representations, without dangerous colored components.
Since there is no proton decay from the gauge sector of trinified theories, our model
is safe from these effects.

\section{$\mathbf{SU(9)\times SU(3)^3}$}

Before concluding, we wish briefly to present an alternative starting point
that can provide a common origin for the GUT-scale equality of gauge couplings
(without the $Z_3$ symmetry) and the existence of the electroweak Higgs doublets.
We consider an $SU(9)\times SU(3)^3$ bulk gauge theory on a $S^{1}/(Z_{2}\times Z_{2}')$
orbifold.  The SU(3)$_C \times $SU(3)$_L \times$ SU(3)$_R$ symmetry of our previous
model is identified with the diagonal subgroup of an SU(3)$^3$ living within SU(9) and the
other SU(3)$^3$ factor, so that
\begin{equation}
\frac{1}{g_{(C,L,R)}^{2}}=\frac{1}{g_{SU(9)}^{2}}+
\frac{1}{g_{(C',L',R')}^{2}} \,.
\label{eq:makesame}
\end{equation}
Here $C'$,$L'$, and $R'$ refer to the three SU(3) factors present before symmetry breaking.  If
these SU(3) gauge groups are somewhat strongly coupled, then Eq.~(\ref{eq:makesame}) leads to
an approximate unified boundary condition for the diagonal subgroup. This
is precisely the idea of ``unification without unification'' described in Ref.~\cite{weiner}.
Note that the bulk gauge symmetry can be thought of as a two-site deconstructed
sixth dimension, with symmetry broken at a boundary.  Generalizations to replicated SU(9) factors
are also interesting, since the primed gauge couplings do not have to be made particularly
large.  In any case, the SU(9) vector multiplet decomposes under the diagonal SU(3)$^3$
subgroup as
\begin{equation}
V_9 : \: \left( \begin{array}{c|c|c} ({\bf 8},{\bf 1},{\bf 1}) &
({\bf 3},{\bf \bar{3}},{\bf 1}) & ({\bf 3},{\bf 1},{\bf \bar{3}}) \\
\hline ({\bf \bar{3}},{\bf 3},{\bf 1}) & ({\bf 1},{\bf 8},{\bf 1}) &
({\bf 1},{\bf 3},{\bf \bar{3}}) \\ \hline ({\bf \bar{3}},{\bf 1},{\bf 3}) &
({\bf 1},{\bf \bar{3}},{\bf 3}) & ({\bf 1},{\bf 1},{\bf 8})
\end{array} \right)\,\,\,.
\end{equation}
We know from ordinary trinification that fields with the quantum numbers of Higgs doublets
live in the $(\mathbf{1,3,\bar{3}})$ representation and its conjugate.  We therefore wish
to find parity assignments that preserve these elements of the chiral adjoint $\Phi$ as well.
With parity transformations defined as in Eq.~(\ref{eq:vphi}), we choose
\begin{eqnarray}
P_{SU(9)}= diag(1,1,1,1,1,1,1,-1,1),\,\,&&
P'_{SU(9)}= diag(1,1,1,-1,-1,1,1,1,1) \,\,\,, \nonumber \\
P_C = diag(1,1,1), \,\,\, P_L &=&diag(1,1,-1), \,\,\, P_R =diag(1,1,-1), \nonumber \\
P'_C = diag(1,1,1), \,\,\, P'_L &=&diag(1,1,1), \,\,\,\,\,\,\, P'_R =diag(1,1,1).
\label{eq:manyones}
\end{eqnarray}
One finds, for example, that the $({\bf 1},{\bf 3},{\bf \bar{3}})$ components of the
SU(9) chiral adjoint $\Phi_9$ has parities
\begin{equation}
\Phi_9({\bf 1},{\bf 3},{\bf \bar{3}}) :\: \left(\begin{array}{ccc}
(-,+) & (+,+) & (-,+)  \\
(-,+) & (+,+) & (-,+) \\
(-,-) & (+,-) & (-,-) \end{array}\right) \,\,\,,
\end{equation}
which indicates the location of one of the Higgs doublets.  Aside from the corresponding
$(+,+)$ entries in the $({\bf 1},{\bf \bar{3}},{\bf 3})$ block, all other components of
$\Phi_9$ have no zero modes.

The orbifold parities in Eq.~(\ref{eq:manyones}) break the SU(9) symmetry to
$SU(8)\times U(1)$ at the $y=0$ fixed point, $SU(7)\times SU(2)\times U(1)$ at $y=\pi R/2$
and to $SU(6)\times SU(2)\times U(1)\times U(1)'$ overall.  The $SU(3)^3$ factors
are broken to $SU(3)_{C}\times SU(2)_{L} \times SU(2)_{R} \times U(1)_L \times
U(1)_R$ overall, but are unbroken at $y=\pi R/2$.  Thus, the most natural way to include
matter fields is by introducing complete ${\bf 27}$'s at the $\pi R/2$ fixed point.

The breaking of the remaining gauge symmetry down to that of the MSSM can be done
with a boundary Higgs sector, as in our previous model.  To determine the necessary
representations, we may pretend the SU(3)$^3$ factor is embedded in another SU(9), and
use the fact that a Higgs $\Sigma \sim ({\bf 9}, {\bf \bar{9}})$ with diagonal vevs will
leave a diagonal SU(9) unbroken.  A straightforward decomposition of
$\Sigma$ in terms of the actual gauge symmetry at $\pi R/2$,
SU(7)$\times$SU(2)$\times$U(1)$\times$SU(3)$^3$, gives the desired representations.  These
break the remaining symmetry down to the diagonal
subgroup SU(3)$_{C}\times $SU(2)$_{L}\times $U(1)$_L \times
$SU(2)$'_{R}\times $U(1)$'_R$.  We may recover the standard model gauge group by including
SU(9) singlet, $\mathbf{(1,3,\bar{3})}$ and $\mathbf{(1,\bar{3},3)}$ boundary Higgs fields, with
the same pattern of vevs found in conventional trinified theories.  Yukawa couplings can
arise via higher dimension operators involving the boundary Higgs fields, and are unsuppressed
in the Higgsless limit, as shown in the previous model; the decoupling of exotic matter fields
also works in the same way.  Color-triplet components of $\Phi_9$ exist, so that
proton decay is not absent, but doublet-triplet spliting is explained naturally via
the orbifold projection.

\section{Conclusions}

We have presented improved models of 5D trinification.  In the first model, unified
symmetry was broken by a combination of orbifold projections and a boundary Higgs sector
that could be decoupled from the theory.  Electroweak Higgs fields appeared economically as
the fifth components of gauge fields. The model demonstrated the existence of a consistent
low-energy theory in which no chiral Higgs fields needed to be added to the theory in an
ad hoc way.  This model is free of proton decay and consistent with gauge unification.  In
the second model, we showed that an additional SU(9) gauge factor could provide a common
origin for the unified boundary condition on the standard model gauge couplings, and
the origin of the electroweak Higgs, via gauge-Higgs unification.  Both models provide new
and explicit realizations of 5D trinified GUTs, and demonstrate a Higgsless approach
that can be applied to other unified theories with rank greater than four.


\begin{acknowledgments}
We thank the NSF for support under Grant No.~PHY-0456525.  We thank Josh Erlich for
useful comments.
\end{acknowledgments}

\begin{thebibliography}{99}

\bibitem{higgsless}
C.~Csaki, C.~Grojean, H.~Murayama, L.~Pilo and J.~Terning,
Phys.\ Rev.\ D {\bf 69}, 055006 (2004)
[arXiv:hep-ph/0305237].

\bibitem{morehiggsless}
G.~Cacciapaglia, C.~Csaki, C.~Grojean and J.~Terning,
Phys.\ Rev.\ D {\bf 71}, 035015 (2005)
[arXiv:hep-ph/0409126];
C.~Schwinn,
Phys.\ Rev.\ D {\bf 71}, 113005 (2005)
[arXiv:hep-ph/0504240];
G.~Cacciapaglia, C.~Csaki, C.~Grojean, M.~Reece and J.~Terning,
arXiv:hep-ph/0505001.

\bibitem{nomura}
Y.~Nomura, D.~R.~Smith and N.~Weiner,
Nucl.\ Phys.\ B {\bf 613}, 147 (2001)
[arXiv:hep-ph/0104041].

\bibitem{carcon}
C.~D.~Carone and J.~M.~Conroy,
Phys.\ Rev.\ D {\bf 70}, 075013 (2004)
[arXiv:hep-ph/0407116].

\bibitem{othertrin}
A.~Demaria and R.~R.~Volkas,
Phys.\ Rev.\ D {\bf 71}, 105011 (2005)
[arXiv:hep-ph/0503224];
C.~D.~Carone,
Phys.\ Rev.\ D {\bf 71}, 075013 (2005)
[arXiv:hep-ph/0503069];
J.~E.~Kim,
Phys.\ Lett.\ B {\bf 591}, 119 (2004)
[arXiv:hep-ph/0403196];
K.~S.~Choi and J.~E.~Kim,
Phys.\ Lett.\ B {\bf 567}, 87 (2003)
[arXiv:hep-ph/0305002];
I.~Gogoladze, Y.~Mimura and S.~Nandi,
Phys.\ Rev.\ D {\bf 69}, 075006 (2004)
[arXiv:hep-ph/0311127].

\bibitem{gh}  See, for example,
I.~Antoniadis,
Phys.\ Lett.\ B {\bf 246}, 377 (1990);
I.~Antoniadis, K.~Benakli and M.~Quiros,
New J.\ Phys.\  {\bf 3}, 20 (2001)
[arXiv:hep-th/0108005];
N.~Haba and Y.~Shimizu,
Phys.\ Rev.\ D {\bf 67}, 095001 (2003)
[Erratum-ibid.\ D {\bf 69}, 059902 (2004)]
[arXiv:hep-ph/0212166];
G.~Burdman and Y.~Nomura,
Nucl.\ Phys.\ B {\bf 656}, 3 (2003)
[arXiv:hep-ph/0210257];
I.~Gogoladze, Y.~Mimura and S.~Nandi,
Phys.\ Lett.\ B {\bf 560}, 204 (2003)
[arXiv:hep-ph/0301014].

\bibitem{gh2}
C.~Csaki, C.~Grojean and H.~Murayama,
Phys.\ Rev.\ D {\bf 67}, 085012 (2003)
[arXiv:hep-ph/0210133];

\bibitem{nomurahall}
L.~J.~Hall, Y.~Nomura and D.~R.~Smith,
Nucl.\ Phys.\ B {\bf 639}, 307 (2002)
[arXiv:hep-ph/0107331].

\bibitem{gregoire}
N.~Arkani-Hamed, T.~Gregoire and J.~Wacker,
JHEP {\bf 0203}, 055 (2002)
[arXiv:hep-th/0101233].

\bibitem{weiner}
N.~Weiner,
arXiv:hep-ph/0106097.

\end{thebibliography}

\end{document}